# Viscoelasticity and cell swirling motion

Ivana Pajic-Lijakovic*, Milan Milivojevic

Faculty of Technology and Metallurgy, Belgrade University, Serbia

iva@tmf.bg.ac.rs

**Abstract**

Although collective cell migration (CCM) is a highly coordinated and fine-tuned migratory mode, instabilities in the form of cell swirling motion (CSM) often occur. The CSM represents a product of the active turbulence obtained at low Reynolds number which has a feedback impact to various processes such as morphogenesis, wound healing, and cancer invasion. The cause of this phenomenon is related to the viscoelasticity of multicellular systems in the context of cell residual stress accumulation. Particular interest of this work is to: (1) emphasize the roles of cell shear and normal residual stress accumulated during CCM in the appearance of CSM and (2) consider the dynamics of CSM from the standpoint of rheology. Inhomogeneous distribution of the cell residual stress leads to a generation the viscoelastic force which acts to suppress CCM and can induces the system rapid stiffening. This force together with the surface tension force and traction force is responsible for the appearance of the CSM. In this work, a review of existing literature about viscoelasticity caused by CCM is given along with assortment of published experimental findings, in order to invite experimentalists to test given theoretical considerations in multicellular systems.

1. Introduction

A more comprehensive account of main features of collective cell migration (CCM) is necessary for deeper understanding of biological processes such as morphogenesis, wound healing, and cancer invasion [1-5]. Ordered trend of cell movement can be locally perturbed which may lead to the cell swirling motion (CSM). However, despite extensive research devoted to study this type of instabilities, especially in 2D systems [6-8], we still do not fully understand the phenomenon. The observed phenomenon represents a part of the elastic turbulence generated under low Reynolds number.

Various soft matter systems such as flexible, long-chain polymer solutions also generate the swirling motion during low Reynolds flow [9,10]. The elastic turbulence, as a consequence of the system viscoelasticity, is accompanied with the rapid stiffening of the system under flow. For the case of polymer solutions, this elastic turbulence is induced by stretching of polymer chains resulting in significant system stiffening which has a feedback impact on the stress relaxation phenomenon. Beside the Reynolds number, additional dimensionless number is formulated i.e. the Weissenberg number $W_i = \frac{v\, \tau_R}{L}$ ($\tau_R$ is the stress relaxation time) for describing the elastic turbulence [10]. The system stiffening can be related to: (1) the stress relaxation phenomenon (polymer solutions [10]) or the residual stress accumulation [4]. It depends on the time scale for: (1) strain change, (2) stress relaxation, and (3) the residual stress accumulation. In contrast to other soft matter systems, the multicellular systems are active, capable of inducing the self-rearrangement which has been treated as an active turbulence [11]. Cell movement leads to the generation of local strains (shear and volumetric) of multicellular system and their long-time changes (a time scale of hours) [4]. These strains induce the generation of corresponding stresses within a multicellular system, their short-time relaxation (a time scale of minutes) and the long-time residual stresses accumulation obtained during many short-time relaxation cycles [4]. Cell normal residual stress is generated within a migrating cell clusters during their movement through crowded environment and during collisions of velocity fronts [4,12-14]. Shear stress can be intensive within the biointerface between migrating cell clusters and surrounding resting cells [14,15]. Inhomogeneous distribution of cell residual stress induces a generation of a viscoelastic force responsible for the system local stiffening which can lead to a CSM [16]. The viscoelastic force is a resistive force and acts always opposite to the direction of cell movement [16].

While the CSM has been observed during CCM of confluent cell monolayers [6,8], swirls formation has not been recognized during a free expansion of monolayers under *in vitro* conditions [12,17]. Lin et al. [7] considered 2D CCM in the confined environment and pointed out that weak local cell polarity alignment (LA) and strong contact inhibition of locomotion (CIL) are the prerequisite for appearance of the CSM. These conditions correspond to a reduced tissue cohesiveness quantified by lower tissue surface tension. Shellard and Mayor [18,19] observed the CSM during 3D CCM of neural crest cells under *in vivo* conditions. Significant attempts have been made to discuss the influence of the processes such as CIL, LA, and epithelial-to-mesenchymal transition (EMT) on the tissue cohesiveness [7,20,21]. However, much less attention has been paid to relate tissue cohesiveness with the cell residual stress accumulation.

The main goals of this consideration are: (1) to emphasize the role of cell shear residual stress and cell normal residual stress in the CSM appearance and (2) to discuss the dynamic of CSM which is related to the mechanical standing waves, from the standpoint of rheology. The cell normal residual stress accumulation induces an increase in cell packing density which can reduce the tissue cohesiveness and this is controlled by the processes such as CIL, LA and EMT [7,22,23]. In contrast, the cell shear residual stress exerts work through the shear stress torque against the tissue cohesiveness and can induce the CSM. This aspect of CCM will be additionally emphasized during this theoretical consideration and discussed on various examples of 2D and 3D CCM.

2. **The role of cell residual stress accumulation in appearance of the CSM**

CCM induces generation of local cell displacement field $\vec{u}_c(r,\tau)$ equal to $\vec{u}_c(r,\tau) = \frac{\sum_i \vec{u}_j \delta(r-r_i)}{n(r,\tau)}$ (where $n(r,\tau)$ is the cell packing density equal to $n(r,\tau) = \sum_i \delta(r-r_i)$, and $\vec{u}_j$ is the displacement field of the j-th cell) and its long-time change, i.e. $\vec{v}_c(r,\tau) = \frac{d\vec{u}_c}{d\tau}$ (where . $\vec{v}_c(r,\tau)$ is the rate of displacement field change). Long-time change of cell displacement field induces a corresponding long-time change of the generated cell strains. These strains induce generation of the corresponding cell normal stress $\tilde{\sigma}_{cV}$ and shear stress $\tilde{\sigma}_{cS}$, their short-time relaxation, and corresponding long-time residual stresses accumulation. Cell stress relaxes through successive short-time relaxation cycles under constant strain per cycle [4,14]. Short-time scale corresponds to a time scale of minutes while a long-time scale corresponds to a time scale of hours.

Viscoelastic force generated during CCM represents the consequence of an inhomogeneous distribution of cell residual stress and was expressed by Pajic-Lijakovic and Milivojevic [16,37] as:

$$\vec{F}_{Tve} = \vec{\nabla} \cdot (\tilde{\sigma}_{Rc} - \tilde{\sigma}_{Rm}) \qquad (1)$$

where $\tilde{\sigma}_{Rc}$ is the cell residual stress equal to $\tilde{\sigma}_{Rc} = \tilde{\sigma}_{RcS} + \tilde{\sigma}_{RcV}$, $\tilde{\sigma}_{RcS}$ is the cell shear residual stress, $\tilde{\sigma}_{RcV}$ is the cell normal residual stress, $\tilde{\sigma}_{Rm}$ is the matrix residual stress equal to $\tilde{\sigma}_{Rm} = \tilde{\sigma}_{RmS} + \tilde{\sigma}_{RmV}$, $\tilde{\sigma}_{RmS}$ is the matrix shear residual stress, and $\tilde{\sigma}_{RmV}$ is the matrix normal residual stress. The cell-matrix interactions which lead to a generation the matrix residual stress should be accounted for consideration of 2D CCM on extracellular matrix (ECM). The viscoelastic force is a resistive force capable of inducing the system local stiffening which reduce and can suppress further cell movement [16]. The stiffening comes from (1) the cell residual stress accumulation and (2) coupling of the cell shear flow with local extension or compression generated during CCM [14,16]. The viscoelastic force is responsible for the appearance the CSM. We would like to postulate a suitable constitutive stress-strain model for the long-time viscoelasticity caused by CCM under confluent conditions and express the cell residual stress, and then discuss the role of cell shear and normal residual stresses in appearance the CSM.

**2.1 A long-time viscoelasticity: the stress-strain constitutive model**

Stress decreases from initial value to the equilibrium value during the relaxation process. Tambe et al. [16], Serra-Picamal et al. [17] and Notbohm et al. [6] estimated the long-time change of residual stress distribution within 2D systems without measuring the stress relaxation phenomenon. Khaligharibi et al. [24] estimated stress relaxation in cell monolayers under externally induced extension. They reported that stress relaxes from *1.8 kPa* to the residual stress value equal to *300 Pa*. The fast relaxation occurs within the first ~*5 s*, followed by a slow relaxation, which reached a plateau after ~*60 s*. Stress distribution within 3D multicellular systems caused by CCM under *in vivo* conditions has not been measured yet. Marmottant et al. [25] considered stress relaxation of various cellular aggregates under constant strain condition caused by the aggregate uni-axial compression between parallel plates. The stress decreased exponentially with the relaxation time equal to 3-14 *min*, while the stress relaxed during 25 min. This time period corresponds to the duration of a single short-time stress relaxation cycle. On the other hand, the strain is constant during the short-time cycle and it only changes from one stress cycle to another. Stress relaxation under constant strain condition represents the characteristic of a viscoelastic solid rather than viscoelastic liquid [26]. In contrast, Tlili et al. [27] considered CCM of MDCK cell monolayer and proposed the Maxwell model as suitable for viscoelastic liquid. The Maxwell model describes stress relaxation under constant strain rate conditions while strain cannot relax [26,28]. Tlili et al. [27] estimated strain changes based on measured distribution of cell velocities while the stress relaxation was not measured. They estimated the so called viscous relaxation time equal to $70 \pm 15\ min$ which represents cumulative effects of cell shape relaxations. This time scale corresponds rather to the strain relaxation time since the stress relaxation time is much shorter. The ability of strain to relax pointed to the viscoelastic solid rather than to viscoelastic liquid behavior. The Maxwell model could be suitable for describing long-time viscoelasticity during CCM of weakly connected cells under larger cell velocities as discussed by Guevorkian et al. [29]. The simplest model for the viscoelastic solid which is capable of describing the stress relaxation phenomenon is the Zener model. This model introduces one relaxation time for stress under constant strain condition and the other relaxation time for strain under constant stress. Pajic-Lijakovic and Milivojevic [14,16] proposed the Zener model for describing local cell stress relaxation as:

$$\tilde{\sigma}_{ci} + \tau_{Ri}\dot{\tilde{\sigma}}_{ci} = E_i\tilde{\varepsilon}_{ci} + \eta_i\dot{\tilde{\varepsilon}}_{ci} \qquad (2)$$

where $\tilde{\sigma}_{ci}$ is the local shear or normal stress ($\tilde{\sigma}_{cS}$ and $\tilde{\sigma}_{cV}$) such that $\tilde{\sigma}_c = \tilde{\sigma}_{ce} + \tilde{\sigma}_{cvis}$, $\tilde{\sigma}_{ce}$ is the reversible –elastic contribution to stress, $\tilde{\sigma}_{cvis}$ is the irreversible –viscous contribution to stress, $\tilde{\varepsilon}_{ci}$ is the corresponding shear or volumetric strain, $E_i$ is the elastic modulus. The Young's modulus equal to $E_1 = \frac{k_B T_{eff}}{V_{eff}}$, $V_{eff}$ is the effective volume per single cell equal to $V_{eff} = \frac{1}{n}$, $n$ is the cell packing density, $k_B$ is Boltzmann constant, $T_{eff}$ is the effective temperature represents a product of cell mobility and has been expressed as $(k_B T_{eff})^{1/2} \sim \langle|\vec{v}_c|\rangle$ (where $\langle|\vec{v}_c|\rangle$ is the cell average speed) [30], while the shear elastic modulus is equal to $E_2 = E_2(E_1, \nu)$ (where $\nu$ is the Poisson's ratio). Volumetric and shear viscosities are equal to $\eta_i = E_i t_{ri}$, $t_{ri}$ is the corresponding strain relaxation time under constant stress condition, $\tau_{Ri}$ is the stress relaxation time for shear and normal stresses, $\dot{\tilde{\sigma}}_{ci} = \frac{d\tilde{\sigma}_{ci}}{dt}$ and $\dot{\tilde{\varepsilon}}_{ci} = \frac{d\tilde{\varepsilon}_{ci}}{d\tau}$. The relaxation of local stress under constant strain $\tilde{\varepsilon}_{ci} = \tilde{\varepsilon}_{ci0}$ (shear or volumetric strain) was:

$$\tilde{\sigma}_{ci}(r,t,\tau) = \tilde{\sigma}_{ci0}\ e^{-\frac{t}{\tau_{Ri}}} + \tilde{\sigma}_{ciR}(r,t_{eq},\tau)\left(1 - e^{-\frac{t}{\tau_{Ri}}}\right) \qquad (3)$$

where $\tilde{\sigma}_{ci0}$ is the initial value of the stress and the residual stress $\tilde{\sigma}_{ciR}$ (shear or normal residual stress) is:

$$\tilde{\sigma}_{ciR}(r, t_{eq}, \tau) = E_i\, \tilde{\varepsilon}_{ci0}(r, \tau) \tag{4}$$

where the local elastic modulus $E_i = E_i(r,\tau)$ and viscosity $\eta_i(r,\tau)$ are constant per single short-time cycle, but can change from cycle to cycle [14]. This correlation between the cell residual stress and the corresponding strain (eq. 4) was confirmed experimentally by Notbohm et al. [6]. They considered CSM within a confluent cell monolayer by measuring a long-time change of stress radial component $\sigma_{crrR}$. They pointed out that the $\frac{d\sigma_{crrR}}{d\tau}$ correlated well with the long-time strain change $\frac{d\varepsilon_{crr}}{d\tau}$. This result directly favors the Zener model as a suitable for describing the viscoelasticity of cell monolayers since it (1) accounts for experimentally obtained correlations between $\frac{d\sigma_{crrR}}{d\tau}$ and $\frac{d\varepsilon_{crr}}{d\tau}$ and (2) describe the stress relaxation [16].

## 2.2 Roles of cell normal and shear residual stresses in the appearance of CSM

Every stress part (normal and shear) has its own defined role in the appearance of CSM which is schematically presented in Figure 1.

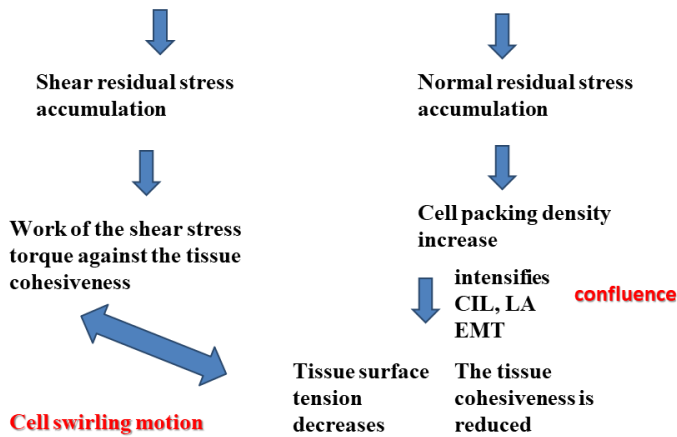

**Figure 1**. The schematic presentation cause-consequence relation between the cell residual stress accumulation and the CSM.

Cell normal residual stress accumulation $\tilde{\sigma}_{cRV}$ induces an increase in cell packing density [12,17,22]. This increase intensifies CIL and reduces LA which leads to a decrease in the tissue surface tension [7]. Mombash et al. [31] measured a decrease in the tissue surface tension during neural aggregate uni-axial compression between parallel plates. The surface tension decreased about 3 times in comparison with the relaxed state. Cell shear residual stress generates work through the shear stress torque against the tissue cohesiveness and thus it can induce the CSM.

The inter-relation between cell normal residual stress and the tissue surface tension has been formulated based on the Young-Laplace equation [25,32]:

$$\tilde{\sigma}_e = \Delta p \, \tilde{I}^j + \tilde{\sigma}_{cRV}{}^j \qquad (5)$$

where $\tilde{\sigma}_e$ is the externally applied stress equal to $\tilde{\sigma}_e = 0$ for the experimental conditions considered here, $\tilde{\sigma}_{cRV}{}^j$ is the cell normal residual stress after the j-th stress relaxation cycle, $\Delta p^j$ is the hydrostatic pressure during the j-th relaxation cycle that is equilibrated with the corresponding value of the tissue surface tension $\gamma^j$ such that $\Delta p^j = \gamma^j H^j$, $H^j$ is the corresponding curvature of the multicellular system part equal to $H^j = \left(\frac{dS}{dV}\right)^j$, $dS(r,\tau)$ is the aggregate surface part, $dV(r,\tau)$ is the corresponding aggregate volume part. Consequently, the change of the isotropic part of cell normal stress represents a consequence of the surface effects and can be expressed as:

$$\frac{d\tilde{\sigma}_{cRV}}{d\tau} = -\frac{d(\gamma H)\tilde{I}}{d\tau} \qquad (6)$$

The normal residual stress accumulation induces a local increase in cell packing density, i.e. $\frac{dn}{d\tau} \sim \tilde{\sigma}_{cRV}$ [22]. The phenomenon can be understood in the context of the plithotaxis [54]. It is further discussed in the **Appendix** 1. The tissue surface tension represents a sum of single-cell contributions, i.e. $\gamma(r,\tau) = \frac{\sum_i \gamma_i \delta(r-r_i)}{n(r,\tau)}$ (where $\gamma_i$ is the single-cell contributions to the tissue surface tension which will be described based on the Voronoi model [7,33]). The normal residual stress equilibrates with the surface tension from cycle to cycle and on that base influences the tissue cohesiveness.

Cell shear residual stress induces generation a shear stress torque equal to $\Delta\vec{T}(r,\tau) = \left(\vec{\nabla}\tilde{\sigma}_{cRS}(r,\tau)\right) X \vec{r}$ (where $\tilde{\sigma}_{cRS}(r,\tau)$ is the cell shear residual stress). The work of shear stress torque $\Delta\vec{T}$ is:

$$\Delta \vec{T} \cdot \vec{\omega}_c \geq \frac{d(\gamma H)}{d\tau} \tag{7}$$

where $\vec{\omega}_c = \vec{\nabla} X \vec{v}_c$ is the angular velocity. Consequently, the accumulation of shear residual stress and lower tissue surface tension accompanied with weaker cell-cell adhesion contacts are prerequisite for the appearance of CSM. The phenomenon is shown schematically in Figure 2 for various 2D multicellular systems.

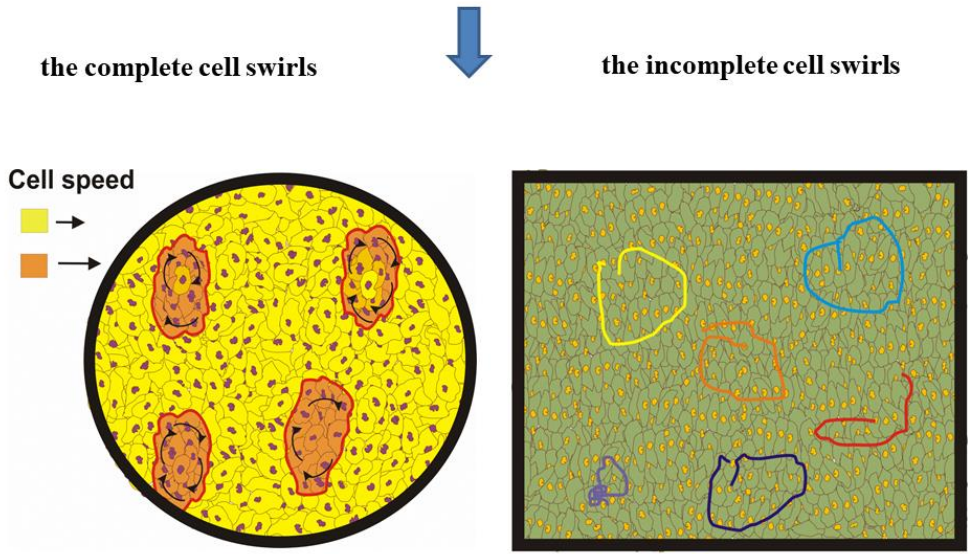

**Figure 2**. The CSM (complete and incomplete) during 2D CCM.

The cohesiveness of MDCK cell monolayer is reduced under confluence and these cells can form complete swirls during CCM [6]. It is in accordance with the fact that the MDCK cells develop weaker cell-cell adhesions [34]. Peyret et al. [35] considered 2D CCM of confluent HaCaT skin cells and Caco2 intestinal cells which form stronger cell-cell adhesion contacts. They reported that these cell lines move in the same direction with no central symmetry, and only the collective direction of motion varies in time by forming circular trajectories in the form of incomplete swirls.

Cells within a swirl undergo complex movement: (1) the successive radial inward and outward flows and (2) azimuthal shear flow as was shown in Figure 3. Successive radial inward flow and outward flow, caused by the action of centrifugal force against to viscoelastic force and the surface tension force, represent the part of the mechanical standing waves well elaborated on the model system such as CCM of a confluent cell monolayer [6,34]. Those mechanical waves represent oscillatory changes of cell velocities and the relevant rheological parameters such as volumetric and shear strains and corresponding residual stresses within a swirl.

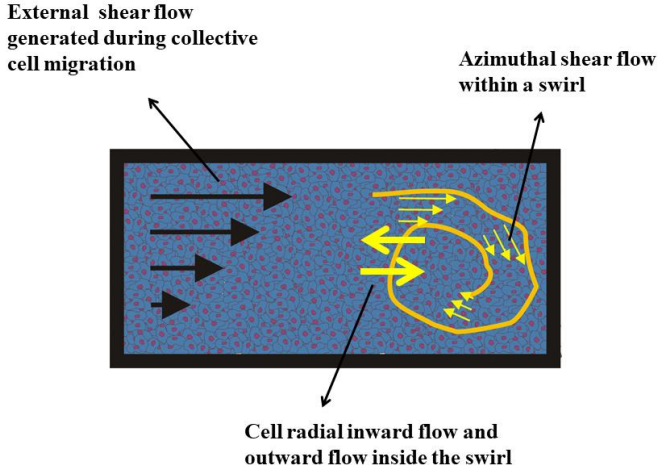

**Figure 3**. Schematic presentation of cell radial inward flow and outward flow and azimuthal shear flow within a swirl together with the shear flow responsible for inducting the CSM.

Azimuthal shear stress component within a swirl $\sigma_{cR\theta r}$ is lower than the macroscopic shear stress component $\sigma_{cRxy}$ which induces the CSM. The epithelial cells are very sensitive to shear stress. Delon et al. [36] reported that shear stress of $3x10^{-3}\ Pa$ induced significant functional changes in Caco2 cell monolayers. The CSM also represents a part of this cell tendency. Detailed description of the mechanical waves generation during CSM will be discussed on 2D cellular system based on postulated force balance [16].

3. The mechanical waves generation during 2D CSM

The CSM has been considered in 2D by using cylindrical coordinates, i.e. $\vec{v}_c = (v_{cr}, v_{c\theta})$ (where $v_{cr} = \frac{dr}{d\tau}$ is the radial velocity component, $v_{c\theta} = r\frac{d\theta}{d\tau}$ is the azimuthal velocity component, and the argument is $\theta = arctg\frac{v_\theta}{v_r}$).

Notbohm et al. [6] discussed cell radial inward flow and outward flow during CSM. They reported that radial component of velocity $v_{cr}$ simultaneously changes a direction every $\sim 4-6\ h$, while the maximum value of the radial velocity was $v_{cr}^{max} \approx 0.25\ \frac{\mu m}{min}$. The velocity $v_{cr}$ is approximately constant within domains $\Delta r \sim 30-40\ \mu m$ during the time period of $\Delta \tau \sim 1-2\ h$ [6]. The radial inward flow and outward flow represent the long-time apparent inertial effects. The main characteristics of standing waves are (1) the radial velocity and cell tractions are uncorrelated, (2) radial stress component $\sigma_{crr}$ and the corresponding strain rate $\dot{\varepsilon}_{crr}$ are uncorrelated, (3) radial stress component is simultaneously tensional and compressional, and (4) time derivative of the stress component is in a phase with the corresponding strain rate [6].

Cell radial outward flow is driven by the centrifugal force against the viscoelastic force and the surface tension force. The work of centrifugal force is responsible for the swirl radial extension such that $v_{cr} > 0$. The centrifugal force decreases with an increase in $r$. Notbohm et al. [6] pointed out that the cell residual stress accumulation is more intensive in the swirl core region for $r < 150\ \mu m$ than in the swirl peripheral region for $150\ \mu m \leq r < 300\ \mu m$. The centrifugal force leads to cell transfer from the region of higher cell velocity near the swirl center to the region of lower velocity which intensifies the CIL.

The viscoelastic force $\vec{F}_{Tve}^{SW}$ generated within a swirl is equal to $\vec{F}_{Tve}^{SW} = \vec{\nabla} \cdot (\tilde{\sigma}_{Rc}^{SW} - \tilde{\sigma}_{Rm}^{SW})$ (where $\tilde{\sigma}_{Rc}^{SW}$ is the cell residual stress within a swirl and $\tilde{\sigma}_{Rm}^{SW}$ is the matrix residual stress generated by CSM). It is a resistive force directed always opposite to the direction of migration. The surface tension force acts to reduce the surface of a multicellular system. It was expressed by Pajic-Lijakovic and Milivojevic [16] as $\vec{F}_{st}^{SW} = \gamma \vec{u}_c^{SW}$. The traction force $\vec{F}_{tr}^{SW}$ acts in the direction of cell movement. It was expressed as $\vec{F}_{tr}^{SW} = k \vec{u}_m^{SW}$ (where $k$ is an elastic constant of cell-matrix adhesion contact and $\vec{u}_m^{SW}$ is the corresponding matrix displacement field exerted by cells) [37].

Cell radial outward flow (for $v_{cr} > 0$), driven by the centrifugal force, induces an intensive coupling between radial elongation flow and azimuthal shear flow and an increase in the viscoelastic force $\vec{F}_{Tve}^{SW}$ which leads to a local system stiffening. Radial elongation induces the local azimuthal compression of a swirl in order to keep the tissue integrity. The radial swirl extension also induces an increase in the surface tension force $\vec{F}_{st}^{SW}$. The viscoelastic force and the surface tension force acts to suppress cell radial outward flow which leads to a decrease in the cell azimuthal velocity component $v_{c\theta}$ as well as a decrease in the centrifugal force. The decrease in the centrifugal force results in cell radial inward flow (for $v_{cr} < 0$) and consequently the local radial compression of a swirl. The radial compression induces (1) a local increase in the cell packing density and the corresponding decrease in the tissue surface tension and (2) an azimuthal extension of swirl parts which results in the swirl local softening. This softening leads to an increase in the velocity component $v_{c\theta}$ and on that base the centrifugal force. Consequently, the viscoelastic (elongation) force resists the outward flow, while the viscoelastic (compressive) force resists the inward flow. The increase in the centrifugal force leads to the system outflow again.

The traction force $\rho_A \vec{F}_{tr}^{SW}$ acts in the direction of cell migration and influences the rate of cell spreading depending on the rheological behavior of a matrix [6,16]. The matrix residual stress $\tilde{\sigma}_{mR}$ accumulation caused by CCM induces the matrix stiffening. Cell movement on the stiffer ECM leads to an increase in the density $\rho_A$ and the reinforcement of the focal adhesions quantified by the elastic constant $k$.

The corresponding 2D force balance for the CSM at a supracellular level was expressed by the modified model proposed by Pajic-Lijakovic and Milivojevic [16]:

For $r$-direction:

$$n(r,\tau)\left(\frac{\partial v_{cr}}{\partial \tau} + v_r \frac{\partial v_{cr}}{\partial r} + \frac{v_{c\theta}}{r}\frac{\partial v_{cr}}{\partial \theta} - \frac{v_{c\theta}^2}{r}\right) = \rho_A F_{tr\,r}^{SW} - F_{Tve\,r}^{SW} - nF_{st\,r}^{SW} \qquad (8)$$

For $\theta$-direction:

$$n(r,\tau)\left(\frac{\partial v_{c\theta}}{\partial \tau} + v_{cr} \frac{\partial v_{c\theta}}{\partial r} + \frac{v_{c\theta}}{r}\frac{\partial v_{c\theta}}{\partial \theta} + \frac{v_{cr}v_{c\theta}}{r}\right) = \rho_A F_{tr\,\theta}^{SW} - F_{Tve\,\theta}^{SW} - nF_{st\,\theta}^{SW}$$

where the internal centrifugal force is equal to $F_C^{SW} = n\frac{v_{c\theta}^2}{r}$ and the force which accounts for coupling between radial elongation (or compression) with azimuthal shear flow is expressed as $F_{CL}^{SW} = n\frac{v_{cr}v_{c\theta}}{r}$. The coupling force $F_{CL}^{SW}$ acts to reinforce the radial flow [9]. The density of cell-matrix adhesion contacts $\rho_A$ can be expressed in the form of Langevin-type equation as:

$$\frac{\partial \rho_A(r,\tau)}{\partial \tau} = \vec{\nabla}\left[D_{eff}\left(\rho_A(r,\tau)\vec{\nabla}\frac{\delta F_A(\rho_A)}{\delta \rho_A(r,\tau)}\right) + \left(\vec{\nabla}\cdot\tilde{\sigma}_{mR}(r,\tau)^{SW} - \vec{\nabla}\cdot\tilde{\sigma}_{cR}(r,\tau)^{SW}\right)\right] \qquad (9)$$

where $D_{eff}$ is the effective diffusion coefficient of moving cell fronts, and $F_A$ is the free energy density of cell-matrix adhesion contacts equal to $F_A = \sum_l \mu_l \delta(r - r_l)$, $\mu_l$ is the chemical potential of the l-th adhesion contact which depends on storage and loss moduli of ECM [38]. The first term on the right-hand side represents the thermodynamic affinity which accounts for cumulative effects of biochemical processes such as cell signalling and gene expression (which guide migration and influence the state of cell-matrix adhesion contacts), while the stress difference represents the mechanical driving force. The CCM induces an accumulation of cell residual stress $\tilde{\sigma}_{cR}(\boldsymbol{r},\boldsymbol{\tau})^{SW}$ and matrix residual stress $\tilde{\sigma}_{mR}(\boldsymbol{r},\boldsymbol{\tau})^{SW}$. The accumulation of $\tilde{\sigma}_{cR}(\boldsymbol{r},\boldsymbol{\tau})^{SW}$ induces an increase in the $n(r,\tau)$ which intensifies the CIL and on that base leads to a decrease in the density $\rho_A(r,\tau)$. In contrary, the accumulation of $\tilde{\sigma}_{mR}(\boldsymbol{r},\boldsymbol{\tau})^{SW}$ induces the matrix stiffening which leads to an increase in the density $\rho_A(r,\tau)$ [39]. An increase in the density $\rho_A(r,\tau)$ leads to the increase in the total cell traction force $\rho_A \vec{F}_{tr}^{SW}$.

The CSM appearance depends primarily on the tissue cohesiveness. In order to deeply understand the phenomenon it is necessary to discuss the processes which influence the tissue cohesiveness.

## 4 Tissue cohesiveness

Radial cell inward flow and outward flow influence local cell packing density and on that base the tissue cohesiveness by intensifying CIL. Intensive CIL suppresses LA and reduces the tissue cohesiveness. EMT, if exist, can induce a mechanical weakening of cells themselves and cell-cell adhesion contacts [2,21]. Alert et al. [23] pointed out that LA dynamics corresponds to a short-time scale (a times scale of minutes) while a change of cell velocities corresponds to the long-time scale (a time scale of hours). Cell polarities establish equilibrium states during the successive short-time relaxation cycles. Every equilibrium state of cell polarities corresponds to current cell configuration and cell velocities.

### 4.1 Force balance equation for CSM at cellular level

Force balance equations at a cellular level have been expressed in the form of the Vertex and Voronoi models [7,20,40]. These models neglect the apparent inertial effects generated during the CSM. However, the apparent inertial effects represent the consequence of cell radial inward flow and outward flow [6]. Consequently, the Voronoi model should be modified by including (1) the apparent inertial effects and (2) cell viscoelasticity in the form of the generalized Langevin equation [41]:

$$m_c \frac{d\vec{v}_{ci}(\tau)}{d\tau} = -\int_0^t \xi(\tau - \tau')\vec{v}_{ci}(\tau')d\tau' - \frac{\partial U}{\partial \vec{r}_i} + \vec{F}_{di} \qquad (10)$$

where $\vec{v}_{ci}(\tau) = \frac{d\vec{r}_i}{d\tau}$ is the velocity of the i-th cell, $\vec{r}_i$ is the position of the i-th cell, $m_c$ is the mass of single-cell, $U$ is the cell potential described in the **Appendix 2**, $\vec{F}_{di}$ is the stochastic driving force for single-cell movement equal to $\vec{F}_{di} = \vec{F}_{Pi} + \vec{F}_{Ri} + \vec{F}_{RANi}$, $\vec{F}_{Pi}$ is the cell self-propulsion force expressed as $\vec{F}_{Pi} = F_m \vec{p}_{eq\,i}$ [20], $F_m$ is the magnitude of the self-propulsion force, $\vec{P}_{eqi}$ is the vector of equilibrium polarity of the i-th cell established after current short-time relaxation cycle $\vec{P}_{eqi} = (cos\varphi_{eqi}, sin\varphi_{eqj})$, $\varphi_{eqj}$ is the equilibrium polarization angle, $\vec{F}_{Ri}$ is the repulsion force associated to the reduction of the cell–matrix adhesion area [20], $\vec{F}_{RANi}$ is the noise term, $\xi(\tau)$ is the memory kernel which represents a delay time distribution [41]. This distribution is the consequence of the mechanical energy dissipation caused by cell-matrix and cell-cell interactions. The memory kernel $\xi(\tau)$ is equal to $\xi(\tau) = \frac{\langle \vec{F}_{di}(\tau+\Delta\tau)\,\vec{F}_{di}(\Delta\tau)\rangle}{\xi_0 k_B T_{eff}}$, $k_B$ is the Boltzmann constant, and $T_{eff}$ is the effective temperature [30].

If the $\vec{F}_{di}$ represents the consequence of the effective thermal fluctuations, it satisfies the conditions for the white noise, i.e. $\langle \vec{F}_{di}(\tau)\rangle = 0$ and $\langle \vec{F}_{di}(\tau + \Delta\tau)\vec{F}_{di}(\Delta\tau)\rangle = \xi_0 k_B T_{eff}\delta(\tau)$ (where $\xi_0$ is the friction coefficient and $\delta(\tau)$ is the delta function). Corresponding memory kernel can be expressed as $\xi(\tau) = \xi_0 \delta(\tau)$ [41]. However, the white noise approximation is very rough. Mechanical standing waves generation during the CSM also influences single-cell fluctuations. Selmetzi et al. [42] reported that the Gaussian white noise is not always a good approximation for modeling the stochastic driving force generated during cell movement. They pointed out that cell migration is affected by sticking and slipping events alternate to produce, a more or less, irregular mode of motion. Dieterich et al. [43] considered movement of MDCK cells and pointed out to the anomalous nature of the cell movement in the form of super-diffusion. The corresponding memory kernel can be expressed as $\xi(\tau) = \xi_0 D_\tau^{-\alpha}$ (where $D_\tau^{-\alpha}$ is the fractional derivative and $\alpha$ is the order of the derivative). For the super-diffusion condition the parameter $\alpha$ is in the range $2 > \alpha > 1$. Caputo's definition of the fractional derivative of a function $f(\tau)$ was used, and it is given as [44]: $D^{-\alpha}f(\tau) = \frac{1}{\Gamma(1+\alpha)}\frac{d}{d\tau}\int_0^\tau \frac{f(\tau')}{(\tau-\tau')^{-\alpha}}d\tau'$ (where $\Gamma(1-\alpha)$ is a gamma function). The super-diffusive nature of CCM was confirmed by Lin et al. [45]. They considered a mesoscale cell turbulence during 2D CCM and revealed that the kinetic energy of migrating cells follows the q-Gaussian distribution rather than the Maxwell–Boltzmann distribution with the q-index higher than 1. Volpe and Wehr [46] considered the diffusion of active particles on ECM by using of the multiplicative noise. The intensity of this noise depends upon the system's state. Deeper insight into the nature of the stochastic driving force $\vec{F}_{di}$ is necessary to formulate the distribution of delay times $\xi(\tau)$.

A short-time LA has been expressed by Langevin equation [20]:

$$\frac{d\varphi_i}{dt} = -f_{CIL}(\varphi_i - \phi_c^{vel}) + \sqrt{2D_R}\,\eta_i^R(t) \qquad (11)$$

where $\varphi_i$ is the polarization angle, $f_{CIL}$ is the cell repolarization rate equal to $f_{CIL} = \frac{1}{t_{pR}}$, $t_{pR}$ is the relaxation time which corresponds to few minutes [23], $D_R$ is the rotation diffusion coefficient, $\phi_c^{vel}$ is the argument of cell velocity equal to $\phi_c^{vel}(r,\tau) = \arg(\vec{v}_c(r,\tau))$ [7]. The relaxation time decreases by increasing the $n(r,\tau)$, i.e. $t_{pR} \sim \frac{1}{n(r,\tau)}$. Increase in $n(r,\tau)$ intensifies the CIL and on that base induces the cell repolarization which suppresses the LA. Cell polarities establish equilibrium states during successive short-time relaxation cycles. The argument $\phi_c^{vel}$ for CSM changes permanently $\phi_c^{vel}(r,\tau)^R = \omega_c\tau$ (where $\vec{\omega}_c = \vec{\nabla}X\vec{v}_c$).

**4.2 Epithelial-to-mesenchymal transition**

Epithelial-to-mesenchymal transition (EMT) is a cellular process during which epithelial cells acquire mesenchymal phenotypes [21]. They pointed out to the main characteristics of EMT: (1) cytoskeleton remodeling, (2) loss of apical–basal cell polarity, (3) cell–cell adhesion weakening, (4) cell–matrix adhesion remodeling, (5) cell individualization, (6) establishment of the front–back cell polarity, (7) acquisition of cell motility, and (8) basement membrane invasion. Mesenchymal cells are softer than the epithelial phenotype. Lekka et al. [47] reported that the Young's modulus of bladder cancer cells is $E = 14.0 \pm 2.2\,kPa$, while for non-malignant ones the modulus is $E = 28.5 \pm 3.9\,kPa$. Weakening of adherens junctions (AJs) can be induced by decreasing the concentration of E-cadherin and in some cases, by increasing in the concentration of N-cadherin [2]. Good example represents a movement of the *Xenopus*' cephalic neural crest cells through confined environment between the epidermis and mesoderm. These cells need to switch their E-cadherin-based AJs to N-cadherin-based junctions.

The EMT is sensitive to the cell residual stress accumulation caused by CCM. Partial (EMT) is provoked by applied compressive stress of *~600 Pa* [48]. Fluid shear stress of only *0.14 Pa* is capable of inducing the EMT in Hep-2 cells [49]. The fluid shear stress of *0.3 Pa* causes the EMT in epithelial ovarian cancer [50]. However, the shear stress accumulated during 2D CCM is much higher. Tambe et al. [12] measured the shear stress caused by cell monolayer free expansion. They reported that the maximum value of shear stress is *~150 Pa*. The shear stress can be significant during 3D CCM, especially within the biointerface between migrating cell clusters and surrounding cells in the resting state [14]. Barriga and Mayor [2] pointed out that the EMT occurs in a border of moving cell cluster.

### 5.CSM caused by 3D CCM

The prerequisites for the appearance of CSM are: (1) cell movement through confined environment and (2) the cell shear residual stress accumulation. 3D CCM under *in vivo* conditions frequently satisfies the first and second conditions. Cell shear stress accumulation could be significant within the biointerface between migrating cell cluster and its surrounding. The surrounding can be made by (1) cells in resting state, (2) slowly moved cells, and (3) ECM. Pajic-Lijakovic and Milivojevic [14] theoretically considered the dynamics at the biointerface between migrating cell cluster and surrounding cells in the resting state. Shellard and Mayor [18,19] discussed the CSM within a neural crest supracell under *in vivo* conditions. The CSM occurs inside the supracell at the biointerface between the rapidly moved cell group (pushed by supracellular contractions of actin cable) and surrounding slowly moved cells. The schematic presentation of the CSM induced during 3D CCM is shown in Figure 4.

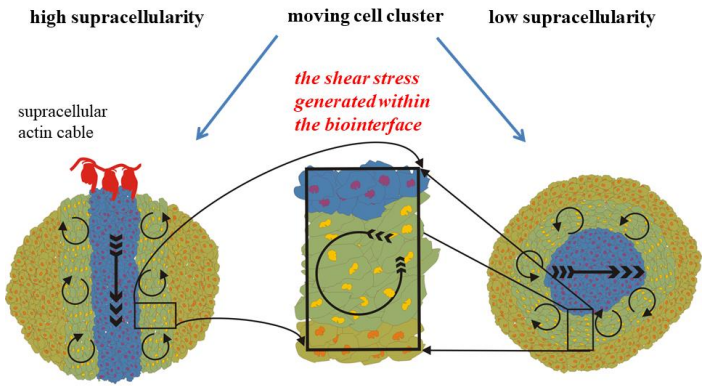

**The cell swirling motion and 3D collective cell migration**

**Figure 4**. The CSM during 3D CCM within the biointerface presented for two moving cell clusters of various levels of supracellularity.

The concept of supracellularity has been used for the group of cells which poses a high level of intracellular organization [18,19]. Consequently, the supracellular unit cannot be treated as a group of single cells. When the cell movement depends mainly on the activities of individual cells, this type of rearrangement corresponds to a low level of supracellularity. In contrast, a high level of supracellularity accounts for the supracellular cytoskeletal organization. Shellard and Mayor [18] discussed the supracellularity based on various factors such as supracellular: (1) polarity, (2) cytoskeletal organization, (3) force transmission, and (4) cell swirling flow. Some of those systems have supracellular actomyosin cable [52].

**5.1 Cell shear stress generation at the biointerface a prerequisite for the appearance of CSM**

Cell shear stress generation depends on slip effects and thickness of the biointerface [15]. Slip effects depend on the velocity gradient generated in the surrounding of migrating clusters. The biointerface can be: (1) sharp for the pronounced slip effects and (2) finite for no-slip effects. Pajic-Lijakovic and Milivojevic ]15] considered influence of the biointerface thickness and slip effects on the viscoelasticity of multicellular system. Cell shear stress consists of viscous and elastic parts, i.e. $\tilde{\sigma}_{cRS} = \tilde{\sigma}_{cRS\,vis} + \tilde{\sigma}_{cRS\,el}$. The average viscous component of shear stress was expressed as [14,15] as:

$$\sigma_{c\ Svis\ xz} \approx \eta \dot{\varepsilon}_{cS\ xz} \quad (12)$$

where $\dot{\varepsilon}_{cS\ xz}$ is the shear rate component equal to $\dot{\varepsilon}_{cS\ xz} = \frac{\Delta v_B}{\Delta L}$, $\Delta L$ is the biointerface thickness, $\Delta v_B$ is the velocity difference across the biointerface equal to $\Delta v_B = v_{Bi} - v_{Bf}$, $v_{Bi} = v_B(z = 0)$ and $v_{Bf} = v_B(z = \Delta L)$ are the velocities at the boundaries of the biointerface, and $\eta$ is the viscosity. Under the no-slip condition, the velocity $v_{Bi}$ is the same as the speed of migrating cell clusters $v_c$. For the intensive slip effects, the finite biointerface becomes the sharp one. Clark and Vignjevic [52] reported that the speed of migrating cell clusters is about $v_c = 0.2 - 1\ \frac{\mu m}{min}$. The biointerface thickness was supposed to be the order of magnitude higher than the size of single-cell, i.e. $\Delta L \approx 100\ \mu m$ [14,15]. The maximum average shear rate is obtained for no-slip condition and $\Delta v_{Bmax} \approx v_c$ is equal to $\dot{\varepsilon}_{cS\ xz}^{max} \sim 0.01\ min^{-1}$. The corresponding viscous part of the residual shear stress is $\sigma_{c\ Svis\ xz}^{max} \sim 15 - 73\ Pa$, calculated for the viscosity $\eta = 4.4 \times 10^5\ Pas$ [25]. The experimental shear residual stress generated during 2D CCM is $100 - 150\ Pa$ [12]. The local azimuthal shear rate for cells within swirl $\dot{\varepsilon}_{cr\theta} = \frac{\Delta v_{c\theta}}{l_c}$ (where $l_c$ is the size of single cell) is significantly lower than the shear rate within the biointerface, i.e. $\dot{\varepsilon}_{cS\ xz} \gg \dot{\varepsilon}_{cr\theta}$ as well as the corresponding shear stress within the swirl. Consequently, the CSM leads to the reduction of shear stress per single cells. The swirls pulsations within the biointerface exert force to the migrating cell cluster and on that base additionally contribute to its movement.

## 6. Conclusion

Significant attempts have been made to describe the main characteristics of 2D CSM which results in the generation of mechanical standing waves. However, the cause of the appearance of this type of instabilities during CCM remains purely understood. The key factor responsible for the CSM is the rapid system stiffening during CCM. The stiffening of multicellular system is induced by the cell residual stress accumulation and its inhomogeneous distribution. The normal residual stress accumulation induces an increase in cell packing density which has a feedback impact on the tissue cohesiveness controlled by the processes such as CIL, LA and EMT. The cell shear residual stress exerts work through the shear stress torque against the tissue cohesiveness and can induce the CSM. The CSM is capable of reducing local shear stress per single cells generated during azimuthal shear flow.

These conditions are satisfied for 2D CCM of confluent MDCK cells. This cell type develops weaker cell-cell adhesions. However, phenotypically and functionally different cells which form stronger cell-cell adhesion contacts, such as HaCaT cells and Caco2 cells, move in the same direction by forming partial circular trajectories. The CSM has been also recognized during 3D CCM. Cell shear stress accumulation during 3D CCM can be significant within the biointerface between migrating cell cluster and its surrounding. The surrounding can be made of (1) cells in resting state, (2) slowly moved cells, and (3) ECM. Two types of the biointerfaces were considered: (1) the biointerface between migrating cluster and surrounding cells in the resting state and (2) the biointerface formed within a neural crest supracell under *in vivo* conditions.

Additional experiments are necessary in order to: (1) correlate generated cell shear stress within a swirl $\tilde{\sigma}_{cRS}^{SW}$ and the shear stress $\tilde{\sigma}_{RcS}$ generated during CCM responsible for inducing the CSM, (2) correlate tissue cohesiveness with the cell packing density, and (3) consider influence of the swirl pulsation on movement of surrounding tissue. This theoretical consideration could help in deeper understanding of various biological processes by which an organism develops its shape and heals wounds in the context of the mechanism supporting the epithelial expansion.

**Acknowledgment**. This work was supported by the Ministry of Education, Science and Technological Development of the Republic of Serbia (No. 451-03-68/2020-14/200135).
**Declaration of interest.** The author reports no conflict of interest.

**Appendix 1: Cell packing density increase as a consequence of normal residual stress accumulation**

Cell normal residual stress accumulation $\tilde{\sigma}_{cRV}$ induces an increase in cell packing density [22]. This increase intensifies CIL and reduces LA which leads to a decrease in the tissue surface tension [7]. A long-time change of cell packing density was expressed by modifying the model proposed by Murray et al. [37]:

$$\frac{\partial n(r,\tau)}{\partial \tau} = -\vec{\nabla} \cdot \vec{J} \tag{A1}$$

where $\vec{J}$ is the flux of cells equal to $\vec{J} = \vec{J}_{conv} + \vec{J}_{cond} + \sum_i \vec{J}_i$, such that $\vec{J}_{conv} = n\vec{v}_c$ is the convective flux, $\vec{v}_c$ is cell velocity (eq. 3), $\vec{J}_{cond} = -D_{eff}\vec{\nabla}n$ is the conductive flux, $D_{eff}$ is the effective diffusion coefficient, $\vec{J}_i = k_i n \vec{\nabla}\phi_i$, are haptotaxis, galvanotaxis, chemotaxis fluxes such that $\phi \equiv \rho$ is the matrix density for the haptotaxis, $\phi \equiv \phi_e$ is the electrostatic potential for the galvanotaxis, $\phi \equiv c$ is the concentration of nutrients for the chemotaxis while $k_i$ are the model parameters which account for various types of interactions such as mechanical, electrostatic or chemical. Durotaxis is the phenomenon of CCM coused by the stiffness gradient of ECM. The durotaxis flux can be expressed as $\vec{J}_d = k_d n \Delta V_m (\vec{\nabla}E_m + \vec{\nabla}G_m)$ (where $k_d$ is the model parameter which represents a measure of matrix mobility induced by cell action, $\Delta V_m$ is the volume of a matrix part, $E_m$ is the matrix Young's modulus, and $G_m$ is the matrix shear modulus). Plithotaxis represents a cell tendency to minimize the inter-cellular shear stress [53]. Steward et al. [54] considered influence of externally applied flow shear stress on generation of the normal inter-cellular stresses. They revealed that even low external shear stress of 1.2 Pa is capable of inducing weakening of cell-cell adhesion contacts and on that base reduction of the inter-cellular normal stress after 12 h. These important experimental findings point to the relation between shear and normal intra-cellular stresses. The plithotaxis is tested at subcellular and cellular levels rather than at the supracellular level [55]. At the supracellular level, the phenomenon can be discussed by inter-relating cell packing density and average normal stress per clusters. Trepat et al. [22] reported that an increase in the average cell normal stress leads to an increase in the $n(r,\tau)$ which has the feedback impact on the cell shear stress generation. Corresponding plithotaxis flux at supracellular level can be expressed as:

$$\vec{J}_p = \mu' n \Delta V_c \vec{\nabla} \cdot \tilde{\sigma}_{cRV} \tag{A2}$$

where $\mu'$ is the mobility of cell velocity front equal to $\mu' = \frac{D}{k_B T_{eff}}$, $D$ is the diffusivity of a velocity front, $T_{eff}$ is the effective temperature, and $k_B$ is the Boltzmann constant, $\Delta V_c$ is the volume of a tissue part, and $\tilde{\sigma}_{cRV}$ is the cell normal residual stress accumulation caused by CCM. The increase in the cell normal residual stress induces an increase in the plithotaxis flux (eq. A2) and on that base an increase in the cell packing density (eq. A1).

**Appendix 2: The cell potential for the Voronoi model**

The potential $U$ consists of active and passive parts $U = U_p + U_a$. The passive potential $U_p$ is $U_p = \sum_i \gamma_i(A_i - A_0) + \sum_{i,j} \Lambda l_{ij}$, $\gamma_i$ is the single-cell contribution to the tissue surface tension equal to $\gamma_i = \frac{1}{2} K_i (A_i - A_0)$, $K_i$ is an effective bulk modulus of the cell, $A_0$ is the reference area of cell while, $A_i$ is the current area of the i-th cell such that $A_i = A_i(\rho_A)$, $\Lambda$ is the adhesion energy per unit length, $l_{ij}$ is the edge length between vertex $i$ and $j$, $U_a$ is the active potential expressed as $U_a = \sum_i \frac{T_{con\,i}}{2} L_i^2$, $T_{con\,i}$ is the contractility coefficient, and $L_i$ is the perimeter of the i-th cell [40].